\documentclass{eptcs}
\providecommand{\event}{FIT 2010} 
\usepackage{breakurl}             
\usepackage[utf8]{inputenc}
\usepackage[english]{babel}
\usepackage{amsmath,amssymb,mathrsfs,stmaryrd,amsthm}
\usepackage{hyperref,xspace}
\usepackage{graphicx,url,color}
\usepackage{alltt}
\usepackage{enumerate}
\usepackage{tabularx}
\usepackage{bbding}
\usepackage[T1]{fontenc}
\usepackage{subfigure}

\newtheorem{definition}{Definition}
\newtheorem{theorem}{Theorem}

\newcommand{\stimes}{\otimes_{\mathit{sy}}}
\newcommand{\atimes}{\otimes_{\mathit{as}}}

\newcommand{\emptyseq}{\epsilon}

\newcommand{\mioI}{\mathit{in}}
\newcommand{\mioO}{\mathit{out}}

\newcommand{\refn}{\leq}
\newcommand{\comp}{\rightleftarrows}
\newcommand{\states}{\mathit{states}}
\newcommand{\start}{\mathit{start}}
\newcommand{\act}{\mathit{act}}

\newcommand{\must}[3]{{\stackrel{#1}{\longrightarrow}\!\!\!{\genfrac{}{}{0pt}{}{#2}{#3}}}}
\newcommand{\may}[3]{{\stackrel{#1}{\dashrightarrow}\!\!\!{\genfrac{}{}{0pt}{}{#2}{#3}}}\,}

\newcommand{\wmr}{\leq_m^*}

\newcommand{\mr}{\leq_m}

\newcommand{\wcomp}{\rightleftarrows_{\mathit{wc}}}
\newcommand{\scomp}{\rightleftarrows_{\mathit{sc}}}
\newcommand{\acomp}{\rightleftarrows_{\mathit{ac}}}

\newcommand{\shared}{\mathit{shared}}

\newcommand{\sd}{\mathcal{A}}

\title{Interface Theories for
  (A)synchronously Communicating Modal I/O-Transition Systems\thanks{This research has been supported by the GLOWA-Danube project 01LW0602A2 sponsored by the German Federal Ministry of Education and Research.}}
\author{Sebastian S.~Bauer \quad \quad Rolf Hennicker \quad \quad Stephan Janisch
\institute{Institut f{\"u}r Informatik\\ Ludwig-Maximilians-Universit{\"a}t M{\"u}nchen, Germany}
\email{\{bauerse,hennicker,janisch\}@pst.ifi.lmu.de}
}

\def\titlerunning{Interface Theories for
  (A)synchronously Communicating Modal I/O-Transition Systems}
\def\authorrunning{S.~S. Bauer, R. Hennicker \& S. Janisch}
\begin{document}
\maketitle

\begin{abstract}
  Interface specifications play an important role in component-based
  software development. An interface theory is a formal framework
  supporting composition, refinement and compatibility of interface specifications.
We present
  different interface theories which use modal I/O-transition systems as their underlying domain
for interface specifications:
 synchronous interface theories, which employ a synchronous
  communication schema, as well as a novel interface
  theory for asynchronous communication where components
  communicate via FIFO-buffers.
\end{abstract}

\section{Introduction}
The idea of an interface theory is to capture
basic requirements that any formalism should obey which is intended
to support the design of components and component systems.
Since system development usually concerns two dimensions,
a horizontal dimension where larger components are built from smaller ones,
and a vertical dimension, where interface specifications are successively refined
(and finally implemented), an interface theory requires concepts of
composition, refinement and compatibility. Of course, it is important that
the different dimensions of system development fit properly together.
Therefore an interface theory requires (at least)  that refinement is preserved by composition
and that compatibility of interfaces is preserved by refinement,
which is needed for independent implementability and reusability of components. 

A formal notion of an \emph{interface theory} was, to our knowledge, first proposed 
by  de Alfaro and Henzinger in~\cite{DBLP:conf/emsoft/AlfaroH01}.
In their work, an interface theory consists of an interface algebra together with a
component algebra thus distinguishing between interface specifications and component implementations.
Later, in \cite{Alfaro2005}, the authors have introduced the term \emph{interface language} which
simplifies the approach by considering just interfaces with the requirements that incremental design and independent implementability
is possible. Interface theory and interface language are abstract concepts which can be instantiated
by concrete formalisms.
The (abstract) notion of an interface theory we shall use hereafter is close to an interface language
but further simplified by concentrating
on the two rudimentary requirements mentioned above which guarantee independent implementability
and which we want to study for particular interface theories supporting synchronous as well
as asynchronous composition.

All interface theories studied in this work use  modal I/O-transition
systems (MIOs),  introduced by Larsen et al.~\cite{larsen2007},~\cite{DBLP:conf/lics/LarsenT88},
as underlying formalism for interface specifications.
MIOs are well suited to describe behavioural properties of reactive components.
They allow to distinguish between transitions which are optional or mandatory for
refinements and thus support loose specification and stepwise development.  
We first
summarize our previous work on interface theories~\cite{tacas2010} which was based on
synchronous composition of MIOs. We discuss strong and weak
versions of refinement and compatibility and we show that both
versions lead to an interface theory.  Then we extend our previous
work and consider asynchronous composition of MIOs which communicate
via output queues.  We introduce the notion of asynchronous
compatibility which requires that each message put in the output queue
of a MIO must eventually be taken by its communication partner
which is related to the requirement of specified reception in communicating
finite state machines~\cite{Brand-Zafiropulo}.  We
show that MIOs with asynchronous composition, asynchronous
compatibility and weak refinement form again an interface theory.
 Finally, we discuss possibilities for verification and further directions
of our work.

\section{Interface Theories for MIOs with Synchronous Composition}\label{sec:synch}

In our study the abstract concept of an interface theory defines rudimentary properties
that should be satisfied by any formal framework for interface specifications.
Given a class $\sd$ of interface specifications, an interface theory
includes a partial composition operator $\otimes$ to combine specifications to larger ones.
The composition operator is, in general, partial
since it is not always syntactically meaningful to compose specifications. 
Interface specifications for which the composition is defined are called composable.
Additionally, an interface theory must offer a refinement relation $\refn$ to
relate ``concrete'' and ``abstract'' specifications, and a compatibility relation
$\comp$ to express when two interface specifications describe components
which can work properly together. In contrast to (syntactic) composability, compatibility
has a semantic flavour related to the behaviour of components. To obtain an interface theory, three requirements must be
satisfied. Obviously, compatible specifications must be syntactically composable.
Moreover, refinement must be compositional in the sense that
it must be preserved by the composition operator and, third, compatibility
must be preserved by refinement.
 
\begin{definition}[Interface Theory]
  An \emph{interface theory} is a tuple $(\sd,
  \otimes, \refn, \comp)$ consisting of a class $\sd$ of interface
  specifications, a partial 
composition operator $\otimes : \sd \times
  \sd \to \sd$,
  a reflexive and transitive refinement relation $\refn\ \subseteq \sd
  \times \sd$, and
  a symmetric compatibility relation $\comp\ \subseteq \sd \times \sd$,
  such that the following conditions are satisfied. Let $S, S', T, T'
  \in \sd$ be interfaces.
  \begin{enumerate}[(1)]
  \item \emph{(Compatibility implies composability)} If $S \comp T$ then $S \otimes T$ is defined.
  \item \emph{(Compositional refinement)} If $S' \refn S$ and $T' \refn T$ and
    $S \otimes T$ is defined, then $S' \otimes T'$ is defined and $S' \otimes T' \refn S \otimes T$.
  \item \emph{(Preservation of compatibility)} If $S \comp T$ and 
    $S' \refn S$ and $T' \refn T$, then $S' \comp T'$.  
  \end{enumerate}
\end{definition}

Obviously, in a top-down design,
the requirements for an interface theory expressed by conditions (1) to (3) support independent development of components
and thus  independent implementability in the sense of~\cite{Alfaro2005}.
To a certain extent an interface theory supports also
bottom-up design, where existing components can be reused as parts of a
larger system architecture, as long as local refinements are correct and local interfaces fit into the context.


In the following we will study particular interface theories which all use modal I/O-transition systems (MIOs)
as their underlying formalism for interface specifications.
Modal I/O-transition systems  have been introduced by Larsen et
al.~\cite{larsen2007},~\cite{DBLP:conf/lics/LarsenT88} as a formalism
to describe the behaviour of reactive, concurrent components.  MIOs
distinguish between 
may- and must-transitions, where the former
model allowed behaviour, which may or may not be present in a refinement,
whereas the latter model required behaviour to be preserved by any refinement.
Thus MIOs support loose specifications and flexible notions of refinement.

\begin{definition}[MIO]
  A \emph{modal I/O-transition system (MIO)} $S =
  (\states_S,\start_S,act_S,\may{}{}{S},\must{}{}{S})$ consists of a
  set of states $\states_S$, an initial state $\start_S \in
  \states_S$, a set $act_S$ of actions being the disjoint union of
  sets $in_S$, $out_S$ and $int_S$ of input, output and internal actions
  resp., a may-transition relation $\may{}{}{S} \subseteq \states_S
  \times act_S \times \states_S$, and a must-transition relation
  $\must{}{}{S} \subseteq \may{}{}{S}$, i.e. every required transition
  is also allowed. The set $act_S$ of actions together with its partition
into input, output and internal actions is called the \emph{signature}
of $S$.
\end{definition}

As usual, we write $s \may{a}{}{S} s'$ instead of $(s,a,s') \in \may{}{}{S}$, and 
similarly for must-transitions.
A state $s \in \states_S$ of $S$ is called \emph{reachable} if
there exist may-transitions $s_0 \may{a_0}{}{S} s_1 \may{a_1}{}{S} \ldots
\may{a_{n-1}}{}{S} s_n$, $n\geq 0$, such that $s_n = s$.
The class of modal I/O-transition systems is denoted by $\mathcal{M}$.
It provides the underlying domain of specifications for all interface
theories considered in the following. 

Two MIOs $S, T \in \mathcal{M}$
are \emph{(syntactically) composable} if their actions only overlap on
complementary types, i.e.  $\act_S \cap \act_T \subseteq (in_S \cap
out_T) \cup (in_T \cap out_S)$. The \emph{set of shared actions} $\act_S \cap
\act_T$ is denoted by $\shared(S,T)$.  The \emph{synchronous
  composition} of two composable MIOs $S$ and $T$ is
defined as the usual product of transition systems with
synchronization on shared actions which become internal in the
product. A synchronization transition in the composition is a
must-transition only if both of the single synchronized transitions
were must-transitions.

\begin{definition}[Synchronous composition]
  Let $S, T \in \mathcal{M}$ be two composable MIOs. The
  \emph{synchronous composition of $S$ and $T$} is the MIO $S \stimes
  T = (\states_S \times \states_T, (\start_S,\start_T), \act,
  \may{}{}{}, \must{}{}{})$ where the action alphabet $\act$ is the
  disjoint union of the input actions $(in_S \cup in_T) \smallsetminus
  \shared(S,T)$, the output actions $(out_S \cup out_T) \smallsetminus
  \shared(S,T)$, and the internal actions $int_S \cup int_T \cup
  \shared(S,T)$. The transition relations are the
  smallest relations satisfying:
\begin{itemize}
\item for all $a \in \shared(S,T)$,
  \begin{itemize}
  \item if $s \may{a}{}{S} s'$ and $t \may{a}{}{T} t'$,
    then $(s,t) \may{a}{}{} (s',t')$,
  \item if $s \must{a}{}{S} s'$ and $t \must{a}{}{T} t'$,
    then $(s,t) \must{a}{}{} (s',t')$,
  \end{itemize}
\item for all $a \in \act_S \smallsetminus \shared(S,T)$,
  \begin{itemize}
  \item if $s \may{a}{}{S} s'$,
    then $(s,t) \may{a}{}{} (s',t)$ for all $t \in \states_T$,
  \item if $s \must{a}{}{S} s'$,
    then $(s,t) \must{a}{}{} (s',t)$ for all $t \in \states_T$,
  \end{itemize}
\item for all $a \in \act_T \smallsetminus \shared(S,T)$,
  \begin{itemize}
  \item if $t \may{a}{}{T} t'$,
    then $(s,t) \may{a}{}{} (s,t')$ for all $s \in \states_S$,
  \item if $t \must{a}{}{T} t'$,
    then $(s,t) \must{a}{}{} (s,t')$ for all $s \in \states_S$.
  \end{itemize}
\end{itemize}
\end{definition}


The basic idea of \emph{modal} refinement is that required
(\emph{must}) transitions of an abstract specification must also occur
in the concrete specification. Conversely, allowed (\emph{may})
transitions of the concrete specification must be allowed by
the abstract specification.
We distinguish between \emph{strong modal refinement},
due to~\cite{DBLP:conf/lics/LarsenT88} and denoted by $\mr$, 
and \emph{weak modal refinement},
due to~\cite{DBLP:conf/ershov/HuttelL89} and denoted by $\wmr$,
which are both defined in terms of a simulation relation.
While in the strong case every transition must be simulated
``immediately'', weak refinement allows to abstract from
transitions with internal actions. We only review the formal definition of
the latter here. In the following, the successive execution of
arbitrarily many internal must-transitions is denoted by
$\must{\tau}{*}{}$, and similarly for may-transitions. 

\begin{definition}[Weak modal refinement] 
    Let $S$ and $T$ be MIOs with the same signature.
    $S$ \emph{weakly modally refines} $T$, written $S \wmr T$,
    if there exists a relation $R \subseteq \states_S \times \states_T$
    containing $(\start_S,\start_T)$ such that
    for all $(s,t) \in R$:\vspace{-0.5mm}
    \begin{enumerate}[(1)]
    \item $\forall a \in in_T \cup out_T : \
      t \must{a}{}{T} t' \Longrightarrow
      \exists\ s \must{\tau}{*}{S} \overline s \must{a}{}{S} \overline{ \overline s}
      \must{\tau}{*}{S} s' \land (s',t') \in R$,\vspace{-1.5mm}
    \item $\forall a \in int_T : \
      t \must{a}{}{T} t' \Longrightarrow
      \exists\ s \must{\tau}{*}{S} s' \land (s',t') \in R$,\vspace{-1.5mm}
    \item $\forall a \in in_S \cup out_T : \
      s \may{a}{}{S} s' \Longrightarrow
      \exists\ t \may{\tau}{*}{T} \overline t \may{a}{}{T} \overline{ \overline t}
      \may{\tau}{*}{T} t' \land (s',t') \in R$,\vspace{-1.5mm}
    \item $\forall a \in int_S : \
      s \may{a}{}{S} s' \Longrightarrow
      \exists\ t \may{\tau}{*}{T} t' \land (s',t') \in R$.
  \end{enumerate}
\end{definition}

In conditions (2) and (4), $a$  is an internal action which must be simulated by
a sequence of arbitrarily many internal actions (denoted by $\must{\tau}{*}{}, \may{\tau}{*}{}$ resp.). This sequence may
be empty but the important point is that the original transition with $a$
must stay in the relation $R$. 

 Our notion of strong modal compatibility
is inspired by~\cite{Alfaro2005} and~\cite{larsen2007}.
Two MIOs $S$ and $T$ are \emph{strongly modally compatible}, denoted by $S \scomp T$,
if they are composable and if for each reachable state $(s,t)$ in the composition $S \stimes T$,
if $S$ \emph{may} send out in state $s$ an action shared with $T$, then
$T$ \emph{must} be able to receive it in state $t$, and conversely.
The difference to \cite{Alfaro2005} and \cite{larsen2007} is that we consider
the ``pessimistic'' case, where MIOs should work properly together in \emph{any}
composable environment while the ``optimistic'' approach, pursued in \cite{Alfaro2005} and \cite{larsen2007}, 
requires the existence of a (helpful) environment; for a discussion see~\cite{DBLP:conf/sigsoft/AlfaroH01}.

Strong modal refinement is compositional w.r.t.\ the synchronous
product~\cite{DBLP:conf/lics/LarsenT88} and preserves strong modal
compatibility~\cite{tacas2010}. Thus we obtain a first interface
theory.  The detailed proof can be found in~\cite{tacas2010_techreport}.
\begin{theorem}\label{thm:strongsync}
  $(\mathcal{M}, \stimes, \mr, \scomp)$ is an interface theory.
\end{theorem}
Weak modal refinement, however, does not preserve strong modal compatibility
due to the possible insertion of internal transitions in the refinement; see~\cite{tacas2010} for a counterexample.
Therefore, we have introduced in~\cite{tacas2010} a weak version of compatibility
such that a communication partner can delay the reception of a message
by performing some internal must-transitions before.

\begin{definition}[Weak modal compatibility]
  Two MIOs $S$ and $T$ are \emph{weakly modally compatible},
  denoted by $S \wcomp T$, if they are composable and if
  for all reachable states $(s,t)$ in $S \stimes T$,
  \begin{enumerate}[(1)]
  \item $\forall a \in out_S \cap in_T : \
      s \may{a}{}{S} s' \Longrightarrow
      \exists\ t \  \must{\tau}{*}{T} \  \overline t \must{a}{}{T} t'$,
    \item $\forall a \in out_T \cap in_S : \
      t \may{a}{}{T} t' \Longrightarrow
      \exists\ s \  \must{\tau}{*}{S} \  \overline s \must{a}{}{S} s'$.  
  \end{enumerate}
\end{definition} 

Since weak modal refinement is compositional w.r.t.\ the synchronous
product~\cite{DBLP:conf/ershov/HuttelL89} and preserves weak modal
compatibility~\cite{tacas2010} we obtain a second interface theory. 
For a detailed proof see again~\cite{tacas2010_techreport}. 
\begin{theorem}\label{thm:weaksync}
  $(\mathcal{M}, \stimes, \wmr, \wcomp)$ is an interface theory.
\end{theorem}

All kinds of refinement and synchronous compatibility notions considered
here are decidable for finite MIOs and can be efficiently computed in time polynomial in the size of
the MIOs.
For further variants of interface
theories with synchronous composition and
for an introduction of the MIO Workbench for refinement and
compatibility checking see~\cite{tacas2010}.

\section{An Interface Theory for MIOs with Asynchronous Composition}

\begin{figure}
  \centering
  \resizebox{0.7\textwidth}{!}{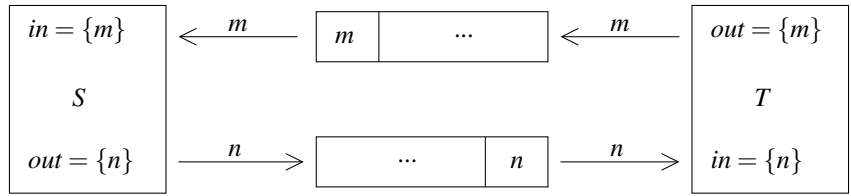 }
  \caption{Asynchronously communicating MIOs}
  \label{fig:async_overview}
\end{figure}

\begin{figure}
  \centering
  \resizebox{0.55\textwidth}{!}{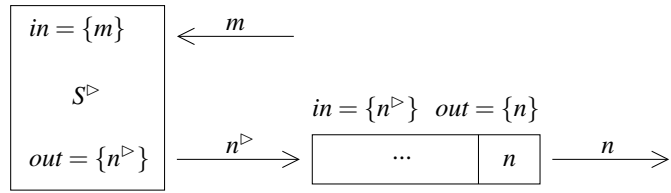 }
  \caption{MIO with output queue}
  \label{fig:mioWithQueue}
\end{figure}

In distributed applications, implemented, for instance, with a message-oriented
middleware, usually an asynchronous communication pattern is used. To
obtain an interface theory for this kind of systems we change the
composition operator and focus on components which communicate via
FIFO-buffered message queues. In Fig.~\ref{fig:async_overview} two
asynchronously communicating MIOs $S$ and $T$ are schematically
depicted: $S$ sends a message $n$ to $T$ by putting it into a queue which stores
the outputs of $S$, and then $T$ can receive $n$ by removing $n$ from
the queue. Obviously, there is a delay between sending and reception.
Similarly, $T$ can send a message $m$ to $S$ by using a second queue which
stores the outputs of $T$. Technically, we enhance MIOs by output
queues which are themselves modelled as MIOs.
Given a MIO $S$ and a distinguished subset $o \subseteq out$ of the output actions of $S$,
the MIO $S$ ``with output queue for the messages in $o$''  is modelled
by the synchronous product of a renamed version of $S$ (where all
 $n \in o$ are renamed to $n^\rhd$) and the ``queue MIO'' $Q_{o}$
which is able to store messages of $o$.
Fig.~\ref{fig:mioWithQueue} shows the idea of this construction where
$S^\rhd$ denotes the renamed version of $S$.

\begin{definition}[MIO with output queue]
Given a set $o$ of output actions, the \emph{queue MIO for $o$} is
$Q_o = (o^*, \emptyseq, \act,
  \may{}{}{}, \must{}{}{})$ where
  the set of states $o^*$ is the set of all finite strings over
  $o$, the initial state $\emptyseq \in o^*$ is the empty string, and
  the set of actions $\act$ is the disjoint union of
  input actions $in = \{ n^\rhd \mid n \in o\}$,
  output actions $out = o$ and with no internal action.
Moreover,
$\may{}{}{} = \must{}{}{}$ and
  the must-transition relation $\must{}{}{}$
  is the smallest relation such that
  \begin{itemize}
  \item for all $n^\rhd \in in$ and states $s \in o^*: \ s \must{n^\rhd}{}{} n\,s$,
  \item for all $n \in out \ (= o)$ and states $s \in o^*: \ s\,n \must{n}{}{} s$.
  \end{itemize}
Given a MIO $S$ with actions $\act_S = in_S \cup out_S \cup
  int_S$ and a distinguished set $o \subseteq out_S$ of output actions,
the \emph{MIO $S$ with output queue for $o$} is given by
  the synchronous product $\Omega_{o}(S) = S_{o}^\rhd \stimes Q_o$
(where $S_{o}^\rhd$ denotes the renamed version of $S$ where all
 $n \in o$ are renamed to $n^\rhd$).
Obviously, the product is well-defined since $S_{o}^\rhd$ and $Q_o$ are composable.
\end{definition}

By the rules of synchronous
composition the input and the output actions of $\Omega_{o}(S)$ coincide
with those of $S$; an output $n$ of $\Omega_{o}(S)$ means that the message
$n$ is either a free output of $S$ or it is removed from the output queue of $S$.  The synchronization actions
$n^\rhd$ of $\Omega_{o}(S)$ express that the message $n$ is put by $S$
(more precisely by $S_{o}^\rhd$) in the queue.

To define the asynchronous composition of two MIOs $S$ and $T$, we assume again
that $S$ and $T$ are composable. Then one can equip $S$ with an output queue
for those outputs $o_S$ of $S$ which can be received by $T$, i.e. which are
shared actions. The other output actions of $S$ remain free.
Similarly $T$ is equipped with an output queue for its shared output actions $o_T$.
Obviously, since $S$ and $T$ are composable, $\Omega_{o_S}(S)$ and $\Omega_{o_T}(T)$
are composable as well. Hence,
two composable MIOs $S$ and $T$ can be asynchronously composed by
synchronously composing their extensions by output queues.

\begin{definition}[Asynchronous composition]
 Let $S, T$  be two composable MIOs and $o_S = out_S \cap in_T$,
 $o_T = out_T \cap in_S$.
  The \emph{asynchronous composition} of $S$ and
  $T$ is defined by $S \atimes T = \Omega_{o_S}(S) \stimes
  \Omega_{o_T}(T)$.
\end{definition}

We consider two composable MIOs $S$ and $T$ to be asynchronously compatible,
if for each reachable state in $S \atimes T$,
if the output queue of $S$ is not empty, then
$T$ must be able to take (i.e.\ input) the next removable element of the queue
possibly after some internal must-transitions, and conversely.
Obviously, due to the use of output queues (instead of input queues),
this idea can be easily formalized with the help of
weak modal compatibility as defined in the synchronous case. 
\begin{definition}[Asynchronous modal compatibility]
  Two MIOs $S$ and $T$ are \emph{asynchronously modally
    compatible}, denoted by $S \acomp T$, if they are composable and if,
for $o_S = out_S \cap in_T$, $o_T = out_T \cap in_S$,
$\Omega_{o_S}(S) \wcomp \Omega_{o_S}(T)$.
\end{definition}
\begin{figure}[htbp!]
  \centering
  \resizebox{0.7\textwidth}{!}{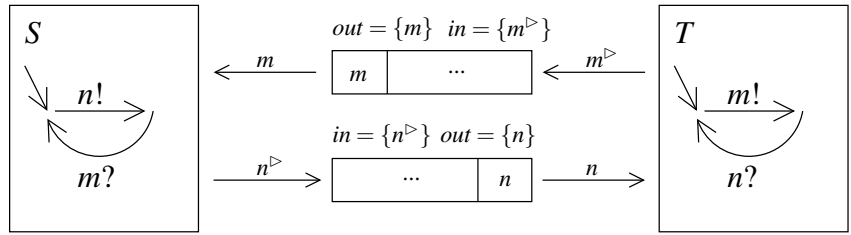 }
  \caption{Example of asynchronously communicating MIOs}
  \label{fig:example}
\end{figure}
As a simple example consider the two MIOs
$S$ and $T$ depicted in Fig.~\ref{fig:example} where
input actions are marked with ``?'' and output actions with ``!'', i.e.
$in_S = out_T = \{m\}$ and $out_S = in_T = \{n\}$.
$S$ has the transitions $\mathit{start}_S \must{n}{}{} s \must{m}{}{}
\mathit{start}_S$, and $T$ has the transitions $\mathit{start}_T
\must{m}{}{} t \must{n}{}{} \mathit{start}_T$. $S$ and $T$ are
asynchronously compatible, since each communication partner must take
the provided message after it has put its own issued message in its
queue (which is an internal must-transition in $\Omega_{o_S}(S)$ and
$\Omega_{o_T}(T)$ resp.). Note that $S$ and $T$ are obviously neither
strongly nor weakly modally compatible which shows the flexibility of
the asynchronous compatibility concept. The other way round it is shown
in~\cite{JanischPHDThesis}
that, under certain conditions like input separated states, weak compatibility implies asynchronous
compatibility.

The behaviour described by the asynchronous composition of MIOs coincides with
the operational model of communicating finite state machines (CFSMs); see~\cite{Brand-Zafiropulo}.
In~\cite{Brand-Zafiropulo} it is required that a system of CFSMs should be well-formed.
One part of the well-formedness condition requires that executable receptions should be specified,
which is just the strong version of the asynchronous compatibility notion used here.
The other direction of the well-formedness condition requires that specified receptions should
be executable. This corresponds to a kind of ``input'' compatibility which we have
not considered here, since, in general, it would not be necessary that any
service offered by a component must actually be used.
Another difference to CFSMs is that we consider a binary (asynchronous)
composition operator but allow open systems, while in the CFSM approach closed
networks of CFSMs are considered.  

To obtain an interface theory with asynchronous composition we still have to choose
an appropriate refinement notion. After a closer look it becomes
obvious that refinement is not really related to the communication paradigm,
since refinement concerns the vertical dimension of software development
moving from abstract to more concrete abstraction levels, whereas
composition is related to the horizontal dimension where larger systems
are constructed from smaller ones and where the underlying communication
schema is crucial. Hence, we can simply reuse the powerful notion
of weak modal refinement which leads to an interface theory
for MIOs with asynchronous composition.
\begin{theorem}\label{thm:async}
  $(\mathcal{M}, \atimes, \wmr, \acomp)$ is an interface theory.
\end{theorem}
\begin{proof}
The proof relies on
the previous results for the synchronous case, since the asynchronous
notions have been defined in terms of the synchronous ones.
  As a first observation, we show that for any two MIOs $S$ and $S'$ and for any
subset $o$ of output actions of $S$ and of $S'$,
    \begin{equation}
      \label{eq:toshowobservation}\tag{I}
S' \wmr S \Longrightarrow \Omega_{o}(S')  \wmr \Omega_{o}(S).
    \end{equation}
Since weak modal refinement is compositional,
by Thm.~\ref{thm:weaksync}, $S' \wmr S$ implies
$S' \stimes Q_{o} \wmr S \stimes Q_o$.
Hence, by definition, 
$\Omega_{o}(S') = S_{o}'^\rhd \stimes Q_o \wmr  S_{o}^\rhd \stimes Q_o = \Omega_{o}(S)$.
  
  We can now prove that the conditions (1) - (3) of an interface theory are satisfied.
  \begin{enumerate}[(1)]
  \item Asynchronously compatible MIOs are, by definition, composable.
  \item Compositionality of refinement: Assume that $S' \wmr S$, $T' \wmr T$
    and that $S \atimes T$ is defined, i.e. $S$ and $T$ are composable. Since
    weak modal refinement $\wmr$ does not change signatures,
$S'$ and $T'$ are composable as well, i.e. $S' \atimes T'$ is defined.

We have to show that
    $S' \atimes T' \wmr S \atimes T$ which means, by definition,
    \begin{equation}
      \label{eq:toshow}\tag{II}
      \Omega_{o_{S'}}(S') \stimes \Omega_{o_{T'}}(T') \wmr
      \Omega_{o_S}(S) \stimes \Omega_{o_T}(T)
    \end{equation}
where   $o_{S'} = out_{S'} \cap in_{T'}$,
$o_{T'} = out_{T'} \cap in_{S'}$, $o_S = out_S \cap in_T$, and
$o_T = out_T \cap in_S$.
First, $S' \wmr S$ implies that $S$ and $S'$ have the same signature;
the same holds for $T$ and $T'$.
Therefore, $o_{S'} = o_S$ and $o_{T'} = o_T$.
    By~(\ref{eq:toshowobservation}), $S' \wmr S$ and $T' \wmr T$ implies
    $\Omega_{o_S}(S') \wmr \Omega_{o_S}(S)$ and $\Omega_{o_T}(T') \wmr \Omega_{o_T}(T)$, respectively.
    Then, (\ref{eq:toshow}) follows from compositionality of
    $\wmr$ w.r.t.\ synchronous composition $\stimes$, see Thm.~\ref{thm:weaksync}, taking into account $o_{S'} = o_S$ and $o_{T'} = o_T$.
  \item Preservation of compatibility under refinement: 
    Assume that $S \acomp T$, $S' \wmr S$ and $T' \wmr T$.
    By definition, $S \acomp T$ means $\Omega_{o_S}(S) \wcomp \Omega_{o_T}(T)$.
From~(\ref{eq:toshowobservation}) we know that
    $S' \wmr S$ implies $\Omega_{o_S}(S') \wmr \Omega_{o_S}(S)$ and
$T' \wmr T$ implies $\Omega_{o_T}(T') \wmr \Omega_{o_T}(T)$.
By Thm.~\ref{thm:weaksync}, $\wcomp$ is preserved
    under $\wmr$ and therefore $\Omega_{o_{S}}(S') \wcomp \Omega_{o_{T}}(T')$.
Thus $\Omega_{o_{S'}}(S') \wcomp \Omega_{o_{T'}}(T')$, since $o_{S'} = o_S$ and $o_{T'} = o_T$ as above.
    This means, by definition, $S' \acomp T'$.
  \end{enumerate}
\end{proof}

\section{Conclusion}
\label{sec:conclusion}
We have studied interface theories based on modal I/O-transition systems (MIOs)
with synchronous and with asynchronous composition. We have chosen MIOs
as the underlying domain for interface specifications since they allow for a flexible refinement notion.
In the synchronous case,
if the underlying MIOs are finite, strong and weak refinement as well as
strong and weak compatibility are decidable and can be efficiently checked with
the MIO Workbench; see~\cite{tacas2010} and~\cite{MIO-Workbench-WWW}. 
In the asynchronous case, the buffering mechanism used for communication may lead to infinite
state spaces.
Concerning refinement it is, however, still possible to derive weak refinements between composed specifications
with infinite state spaces, say $S' \atimes T' \wmr S \atimes T$, from
local refinements $S' \wmr S$ and $T' \wmr T$ and the latter can be decided if the
local MIOs are finite. This is an important consequence of the interface theory
with asynchronous composition.
The situation is different, if we consider the verification of asynchronous compatibility
which is, in general, not decidable due to the potentially infinite output queues.
We are currently working on criteria for asynchronous compatibility,
which are decidable and powerful at the same time, and on the integration of such criteria
into the MIO Workbench. As an outcome of our theoretical work, we want to apply
the results to provide a solid basis for modelling hierarchical and asynchronously communicating components
in the context of the Unified Modeling Language (UML).
At the same time we are also interested in interface theories
for components with local data states~\cite{facs2009, sbmf2010} and for timed systems.

\paragraph{Acknowledgement.}
An important input for this study was the suggestion of Alexander Knapp
to use output queues (instead of input queues) for the formalization of asynchronous
compatibility. We are grateful to Alexander for this very valuable hint.

\vspace{-4.3mm}
\bibliographystyle{eptcs}
\bibliography{bibliography}

\end{document}